\documentstyle[11pt,aaspp4]{article}
\begin{document}

\title{The \it Ulysses \rm Supplement to the BATSE 3B Catalog of Cosmic
Gamma-Ray Bursts}
\author{K. Hurley}
\affil{University of California, Berkeley, Space Sciences Laboratory,
Berkeley, CA 94720-7450}
\authoremail{khurley@sunspot.ssl.berkeley.edu}
\author{M. S. Briggs, R. M. Kippen}
\affil{University of Alabama in Huntsville, Huntsville AL 35899}
\author{C. Kouveliotou\altaffilmark{1}, C. Meegan, G. Fishman} 
\affil{NASA Marshall Space Flight Center, Huntsville AL 35812}
\author{T. Cline}
\affil{NASA Goddard Space Flight Center, Code 661, Greenbelt, MD 20771}
\author{M. Boer}
\affil{Centre d'Etude Spatiale des Rayonnements, B.P. 4346, 31029
Toulouse, France}

\altaffiltext{1}{Universities Space Research Association, Marshall Space
Flight Center ES-84, Huntsville, AL 35812}

\begin{abstract}

We present Interplanetary Network localization information for 218 gamma-ray bursts
in the 3rd BATSE catalog, obtained by analyzing the arrival times of
these bursts at the \it Ulysses \rm and \it Compton Gamma-Ray Observatory \rm (CGRO) spacecraft.  
For any given
burst observed by these two spacecraft, arrival time analysis
(or ``triangulation'') results in an annulus of possible arrival
directions whose half-width varies between 7 arcseconds and 32 arcminutes, depending
on the intensity and time history of the burst, and the distance of the \it Ulysses \rm
spacecraft from Earth.  This annulus generally
intersects the BATSE error circle, resulting in an average reduction of the
error box area of a factor of 30.

\end{abstract}


\section{Introduction}

A knowledge of the precise positions of cosmic gamma--ray bursts (GRBs)
is important for many studies.  For example, recent observations of the optical counterparts to 
bursts have provided evidence that at least two of them are at cosmological
distances (Metzger et al. 1997, Kulkarni et al. 1998) and that, by implication, most or all of
them may also be.  One optical counterpart has been associated
with a galaxy, and the majority of the models for the energy
release in bursts have long invoked neutron star-neutron star or neutron star-
black hole mergers (e.g. Paczynski 1991).  If this is correct, GRBs should originate in or close
to galaxies, and correlations between burst positions and the large scale structure 
of the universe might be expected.  In this context, it has been debated whether
the positions of bursts are correlated with those of Abell clusters (Kolatt \& Piran 1996; 
Hurley et al. 1997a; Struble \& Rood 1997) and radio-quiet quasars (Hurley et al. 1997b;
Schartel, Andernach, and Greiner, 1997).  In both cases, increased precision in the
location accuracy leads to more stringent tests.  Similarly, precise positions greatly improve
searches for quiescent counterparts to GRBs in other wavelength ranges (e.g. Tokanai et al. 1997),
and can help confirm proposed associations between GRBs and fading counterparts (Hurley et al.
1997c).  Finally, burst recurrence from a single source is excluded in most cosmological models.  
Such recurrence would reveal itself through small-scale clustering in the angular distribution
of burst locations, and the sensitivity of the distribution function to repeating sources is
greater when the burst location accuracy is improved (Kippen, Hurley, \& Pendleton 1998). 

We present here a catalog of 218 gamma--ray bursts from the detector aboard the \it Ulysses \rm
spacecraft, which were also observed by the Burst and Transient Source Experiment
(BATSE), and cataloged in the 3rd BATSE catalog (Meegan et al. 1996).  In the
present catalog, their positions have been improved in accuracy by arrival time analysis, or
``triangulation''.  Detection of an event by these two widely
separated instruments leads to a narrow annulus which intersects the BATSE
error circle and reduces its area
by a factor of $\approx$ 30.  Such annulus/error circle intersections have
played important roles in the examples cited above.

\section{Instrumentation}

The \it Ulysses \rm GRB detector (Hurley et al. 1992) is one instrument in the 3rd
Interplanetary Network of burst detectors.  It consists of two 3 mm thick
hemispherical CsI scintillators
with a projected area of about 20 cm$^2$ in any direction.  The detector is mounted
on a magnetometer boom far from the body of the spacecraft, and therefore has a practically
unobstructed view of the full sky.  Because the \it Ulysses \rm mission is in interplanetary
space, the instrument also benefits from an exceptionally stable background.  The
GRB detector operates continuously, and over 97\% of the data are recovered.

A unique feature of the mission is the fact that it is in heliocentric
orbit with aphelion $\approx$ 5 AU.  Thus it reaches an apogee of $\approx$
6 AU or 3000 light-seconds, and during portions of the orbit, the
GRB experiment is farther than any burst detector
has ever been from Earth, resulting in greatly improved location accuracies
for gamma-ray bursts.

The GRB instrument observes bursts in both a triggered and a real-time mode.
The triggered mode is initiated when the number of counts in one of two possible
time intervals exceeds a preset value.  Because the background is stable, this
number is defined in terms of an absolute number of counts, rather
than a number of standard deviations above the background.  The two time intervals
may be selected among values from 0.125 to 4 s.  Each of the two time intervals is associated
with an independently selectable energy window, whose lower and upper thresholds lie
between $\approx$ 15 and 150 keV.  The data collected in the triggered mode consist
of energy spectra, and either 16 s of 8 ms count rates, or 64 s of 32 ms count rates, in the $\approx$
25 to 150 keV energy range; $\approx$ 1.8 and 7.4 s of the data stream record the
time histories prior to trigger, respectively.  Following a trigger, data are
read out for $\approx$ 40 m, during which time the experiment cannot re-trigger.
An example is shown in Figure 1.

Independent of the trigger, a continuous stream of real time data is transmitted.
This consists of count rates taken with 0.25, 0.5, 1, or 2 s time resolution.  The
resolution depends on the spacecraft telemetry rate, and the energy range is $\approx$
25 to 150 keV, as for the trigger.  Data are compressed in this mode.  An example
is shown in Figure 2.
The real time data serve many purposes.  For
example, numerous gamma-ray bursts which are not intense
enough to trigger the
detector can be reliably identified in the real time data, as we demonstrate below.
Bursts arriving during trigger readout may also be identified.
Finally, bursts whose duration exceeds the duration of the triggered memory can
be fully recorded in this mode.  

BATSE consists of eight detector modules situated at the corners of the \it Compton
Gamma-Ray Observatory \rm spacecraft.  Each contains a Large Area Detector (LAD), a 
50.8 cm diameter by 1.27 cm thick
NaI scintillator.  The trigger algorithm examines the count rates of two or more
detectors in the 50 - 300 keV energy range, over 64, 256, and 1024 ms intervals.  The
experiment is described in more detail in Meegan et al. (1996).  For most of the
events in this catalog, we have utilized the DISCSC and PREB data types summed over
two, three, or four detectors.  This gives a 0.064 s resolution time history from 2 seconds prior to
the trigger to 240 s or more after the trigger.  We select discriminator channels 1
and 2, corresponding to the energy range 25 - 100 keV, to match the \it Ulysses \rm
energy range as closely as possible.  Rarely, this data type is not available; in that
case we use MER, TTS, TTE, or 1 s resolution data.  As BATSE is in low Earth
orbit, the time history data are occasionally interrupted by data gaps and Earth occultation.

\section{Search Procedure}

Every cosmic burst detected by BATSE is systematically searched for in the
\it Ulysses \rm data as soon as the BATSE data are available for it.
This is done by using the approximate arrival direction
from BATSE and the position of the \it Ulysses \rm spacecraft to calculate
a range of possible arrival times at \it Ulysses \rm.  The calculation assumes generous uncertainties
in the arrival direction to assure that the crossing window always includes the
burst.  Typical window lengths are 300 - 500 s (e.g., Figure 2).  
In cases where the BATSE positions undergo substantial revision, a
revised search window is calculated and the search is redone.

To carry out the search, the \it Ulysses \rm real time count rates are extracted for
the crossing window and plotted.  The data are searched both automatically
and by eye.  In the automatic procedure, the background count rate is typically
calculated over ten time intervals, and a $>4 \sigma$ increase is searched
for over 2 or more consecutive time intervals.  The search by eye confirms the
presence of the increase, and in some cases, identifies increases that were
not detected in the automatic procedure due, for example, to a long rise time.   
If the burst is intense enough to trigger the \it Ulysses \rm GRB detector,
burst detection is usually immediately apparent in both the real time and
the triggered data streams, with one noteworthy exception.  Very short bursts 
(durations $<$ 0.25 s) can trigger, but remain difficult to identify in the
lower time resolution real time data.  Conversely, long bursts which do not
reach large peak intensities can be identified in the real time data, but may
not trigger regardless of their time-integrated fluxes.

\section{Deriving Annuli by Comparing BATSE and \it Ulysses \rm \bf Time Histories}

When a GRB arrives at two spacecraft with a delay $\rm \delta$T, it may be
localized to an annulus whose half-angle $\rm \theta$ with respect to the
vector joining the two spacecraft is given by cos$\rm \theta$=c$\rm \delta$T/D,
where c is the speed of light and D is the distance between the two
spacecraft.  (This assumes that the burst is a plane wave, i.e. that its
distance is much greater than D.)  The annulus width d$\rm \theta$, and thus one dimension of
the resulting error box, is c $\rm \sigma(\delta$T)/Dsin$\rm \theta$ where
$\rm \sigma(\delta$T) is the uncertainty in the time delay.  This term has
two components, one statistical and one systematic.  The first is
associated with the uncertainty in comparing the time histories of a burst
recorded by two different instruments; the second is due to the uncertainty
in the clock calibrations of the two spacecraft.  A third source of error
is the uncertainty in the knowledge of the spacecraft positions.  We
discuss each in turn.

In the simple case of identical detectors with identical backgrounds,
the most probable delay $\rm \delta$T and its error $\rm \sigma(\delta$T) may
be estimated as follows.  Denote the two background-subtracted time
histories by X$_i$, Y$_i$, i=1,...m, that is, X$_1$ is the number of
counts in the first time bin for detector X, and Y$_1$ the number of
counts in the first time bin for detector Y, and so on.  Assume that the
background can be estimated with arbitrary accuracy, and further that each value of X$_i$ and
Y$_i$ is an independent normal random variable with
variance X$_i$ and Y$_i$ respectively.  For simplicity of notation, assume
that the two time histories recorded by detectors X and Y are perfectly
aligned when time bin 1 of detector X corresponds to time bin 1 of detector Y.
Then for this perfect alignment, corresponding to zero lag between the two
time histories, the quantity $R_i=(X_i-Y_i)/ \sqrt{X_i+Y_i}$
 is a random
variable with mean zero and variance unity for every i.  Thus
\begin{equation}
\sum_{i=1}^{n}R_i^2
\end{equation}
is distributed as $\chi^2$ with n degrees of freedom.  Here the sum
is performed over some number n of time bins, where n $\le$ m.  Now allow the lag
to vary slightly about zero, so that time bin i of detector X corresponds to time bin
j of detector Y, and form the random variable $R_{ij}=(X_i-Y_j)/ \sqrt{X_i+Y_j}$
where the lag ij is now variable.
The quantity
\begin{equation}
r_{ij}^2=\sum R_{ij}^2
\end{equation}
where the summation is over the intervals which X and Y have in common, 
is similarly distributed as $\chi^2$.  In practice we do not
know \it a priori \rm the lag at which the best correlation will occur, so we test
various lags ij until r$_{ij}^2$ reaches a minimum, r$_{min}^2$.  Then the lag or
time delay corresponding to a 3$\sigma$ equivalent confidence level is 
r$_{ij}^2$=r$_{min}^2$+9.  (All annuli in this catalog have widths corresponding
to 3$\sigma$ equivalent confidence.)

Although this simple example describes the essential features of the method, there
are numerous complicating factors in practice.  For BATSE, the background varies
over the orbit, and its value may not always be known accurately; the
assumption of a normal distribution may also be violated when Cyg X-1 is
in the field of view.  The BATSE and \it Ulysses \rm detectors
have different areas, burst and background count rates, and
time resolutions.  We treat this by first, normalizing the BATSE background-
subtracted time history to the \it Ulysses \rm one so that two time histories have the
same total number of counts.  We also generate a BATSE time history with identical
time resolution to the \it Ulysses \rm one by scaling the counts in the original BATSE
time bins.  These adjustments complicate the above expressions
considerably.  However, through Monte-Carlo simulations, as well as time history
comparisons using a completely different method (Laros et al. 1998), we have
verified that the procedure gives correct results.    In
addition to calculating the r$_{ij}^2$ statistic, we also compute the
correlation coefficient as a function of lag, namely
\[
\rho_{ij}=\frac{ \sum(X_i-\langle X \rangle)(Y_j-\langle Y \rangle)}
{\left\{ \sum(X_i-\langle X \rangle)^2 \sum(Y_i-\langle Y \rangle)^2)
\right\}^{0.5}}
\]
where the summations are over the intervals that X and Y have in common,
and find its maximum value.
The purpose of this calculation is to assure that the r$_{ij}^2$ statistic has
indeed identified the best lag; we require that the maximum $\rho_{ij}$ and the
minimum r$_{ij}^2$ correspond to identical lags, or at worst, that the difference
between the lags be much
less than the 3$\sigma$ confidence interval.  Finally, parabolic fits are done to
both statistics to identify the best lag and the confidence intervals.  An example
is shown in Figures 3 and 4.

The second uncertainty, due to the clock calibrations, is in principle much
smaller than the 3$\sigma$ confidence interval.  The clock on the \it Compton
Gamma-Ray Observatory \rm is accurate to 100 $\mu$s, and this accuracy is verified
through pulsar timing.  Onboard software increases the uncertainty in the BATSE trigger times to
$\approx$ 1 ms.  The clock on the \it Ulysses \rm spacecraft is calibrated during
each daily tracking pass, and the clock drift is extrapolated until the next pass to obtain the
correct time; the calibration is verified at
six month intervals by sending
commands to the GRB experiment at precisely known times.  The exact procedure
is described in Hurley (1994).  Due to spacecraft buffering of the commands, the
result is that \it if \rm it can be assumed that no errors larger than 125 ms
are present, \it then \rm the clock can be shown to be accurate to several
milliseconds.  For the purposes of triangulation, however, we take the
extremely conservative approach that no 3 $\sigma$ cross-correlation uncertainty
is less than 125 ms.  Larger timing errors have been identified in the past for both missions.
The \it CGRO \rm clock was found to have a consistent 2 s offset in the early
part of the mission, and random timing errors between 0.25 and several hundred
seconds occurred in certain types of \it Ulysses \rm data.  In all cases the
data were reprocessed to obtain the results in this paper.

Finally, the uncertainties in the spacecraft ephemerides are truly negligible.
The \it Ulysses \rm range is known to $\approx$ 10 km, and the right ascension and
declination of the spacecraft are accurate to $\approx$ 0.00003 degrees.  The
GRO range is known to better than 50 km, and the right ascension and declination
to better than 0.01 degrees.  The error that these uncertainties produce in
the knowledge of the interspacecraft vector (i.e., the center of the annulus) is
of order 0.0005 degrees, and therefore usually two orders of magnitude smaller
than the timing uncertainties.  As a conservative measure, this uncertainty is added
linearly to the uncertainties in the annulus parameters caused by the timing.  It
typically increases the annulus width by several tenths of an arcsecond.

The radius of each annulus and its right ascension and declination are
transformed to a heliocentric frame.  This is equivalent to an aberration
correction (maximum value $\approx$ 20 arcseconds), and involves two adjustments.  
The first is to the spacecraft time.
The signal from a distant spacecraft may travel several thousand seconds on its
way to Earth.  Depending on the Earth's velocity vector relative to the spacecraft,
the relative motion of the Earth and the spacecraft may not be negligible
during this period.  ``Geocentric'' clock times are
related to the position of the Earth when the signal is \it received \rm from the
spacecraft.  ``Heliocentric'' clock times, i.e., times related to the position
of the Earth when the signal was \it transmitted, \rm must be used.  The
difference between the two may be
as large as several hundred milliseconds, and is often comparable to or
greater than the statistical uncertainty of the time history comparison.
The second adjustment is to the inter-spacecraft vector.  Two observers moving
with the Earth and with the sun see different vectors (i.e., annulus centers)
as a result of their motions during the burst's transit time between the
two spacecraft.  This correction, too, may not be negligible.  The result of
triangulating the 1993 January 31 burst is shown in Figure 5.

Although there have never been more than three widely separated spacecraft
in the 3rd IPN, which would have provided redundant determinations and
therefore a verification of burst
positions, there have been numerous instances where a burst source with
a known position was triangulated with \it Ulysses \rm, BATSE, and other
spacecraft.  These include the soft gamma repeater SGR1806-20 and the bursting
pulsar GRO J1744-28 (Hurley et al. 1996, 1998a), the 1997 January 11 burst localized by
BeppoSAX (Galama et al. 1997), the 1997 February 28 burst, with both a BeppoSAX location and an
optical transient (Hurley et al. 1997c), and numerous other BeppoSAX bursts (e.g. Hurley et al.
1998b).  In each case the triangulated position was in good agreement with the known
position of the source.

\section{Burst Selection Criteria}

We have used several selection criteria for the bursts in this catalog.
The first is that the burst must have been detected by \it Ulysses \rm and
BATSE.  Because of the very different detector sizes, all the bursts satisfying
this criterion are observed to be rather intense by BATSE, and therefore
detection by BATSE is unambiguous.  However, the weaker events (as observed
by \it Ulysses \rm in the real time mode) may be difficult to identify.  In these cases,
we rely on several supplementary semi-empirical criteria.  One is that the  r$_{min}^2$ statistic,
divided by the number of degrees of freedom, be less than unity.  Another is
that the maximum value of the correlation coefficient $\rho$ be greater than
0.5.  Yet another is that the ratio of \it Ulysses \rm counts to BATSE counts lie
in the range 0.019 $\pm$ 0.009.  It might seem unusual that the average value of
this ratio is 0.019, whereas the ratio of the the \it Ulysses \rm
projected area in any direction to a single BATSE LAD area is closer to 0.01, 
and the BATSE counts are derived from at least two LAD's in this procedure.  
Several factors account
for this.  First, the \it Ulysses \rm energy range is wider than BATSE's:
25 - 150 keV \it vs. \rm 25 - 100 keV.  Second, the \it Ulysses \rm lower
energy threshold actually allows photons with energies as low as 15 keV to
be counted, albeit with lower efficiency.  Third, the two \it Ulysses \rm hemispherical
detectors are thin, so photons which do not interact in one hemisphere may
interact in the other, increasing the effective area.  Finally, the BATSE
count rates are not corrected for dead time, which is known to affect the
brighter bursts (which are in general among the bursts we observe).  

This set of criteria is not an absolute one.  To a good approximation, every
gamma-ray burst is different from every other one.  In addition, data gaps,
Earth occultation, 
telemetry noise, unusual background behavior, and different
data modes complicate some analyses.  In some cases, bursts
satisfying all the criteria above are rejected because they are simply too
weak to allow reliable triangulation.  In other cases, bursts which do not
satisfy one or more criteria may be accepted.

A final criterion for inclusion in this catalog is that the burst must
not have been observed by a third interplanetary spacecraft.  The \it Ulysses \rm/
BATSE bursts also observed by \it Mars Observer \rm have been published in Laros et al.
(1997).  The \it Ulysses \rm/BATSE bursts also observed by \it Pioneer Venus Orbiter \rm are
in press (Laros et al. 1998).

\section{A Few Statistics}

Over the period covered by the 3B catalog, \it Ulysses \rm observed 346 confirmed
cosmic gamma-ray bursts (i.e., bursts that were observed by at least one
other spacecraft).  Of these, 274 were observed by \it Ulysses \rm and BATSE and
in some cases, other spacecraft as well.  \footnote{A list of all cosmic bursts and the spacecraft which detected them may be found
at http://ssl.berkeley.edu/ipn3/index.html.}
Since BATSE observed 1122 bursts during this period, \it Ulysses \rm observes
approximately one out of every 4 BATSE bursts.
Of the 274, 64 were observed by 
\it Ulysses \rm in triggered mode, and 210 in real time mode.  Finally, 220 were
observed by \it Ulysses \rm and BATSE, but not by a third interplanetary spacecraft;
two of these could not be triangulated according to the criteria explained
above.  

The histogram of Figure 6 shows the distribution of annulus half-widths for the
218 bursts localized.  The smallest is about 7 \arcsec, the
largest 32 \arcmin, and the average is 4.5 \arcmin.  158 of the annuli,
or 72\%, intersect the BATSE 1 $\sigma$ error circles, whose radii are defined by 
$\rm r_{1\sigma}=\sqrt{\sigma_{stat}^2 + \sigma_{sys}^2}$,
where $\sigma_{sys}$ is the systematic error, 1.6$^{\rm o}$,
and $\sigma_{stat}$ is the statistical error.  Figure
5 shows one example.  This is less than
the number which would be predicted (87\%).  An analysis of a preliminary
version of the IPN catalog describes several more complicated BATSE error
models that are consistent with the BATSE-IPN separations (Briggs et al. 1998a).
A more extensive analysis, utilizing the Ulysses supplement to the
4B catalog, is in progress (Briggs et al. 1998b).
One quantity of interest is how close the
annulus passes to the center of the error circle.  Let 
$\alpha_1, \delta_1$ be the right ascension and declination of
the center of a BATSE error circle, and let $\alpha_2, \delta_2, \theta_2$
be the right ascension, declination, and radius of an annulus.
Then the minimum distance between the error circle and the annulus
is given by
$d=\mid \theta_2 - \cos^{-1}(\sin(\delta_1) \sin(\delta_2) +
\cos(\delta_1) \cos(\delta_2) \sin(\alpha_1 - \alpha_2) ) \mid $.
A histogram of the distribution of minimum distances
between the annuli and the centers of the BATSE error circles
is given in Figure 7.

In general, the annuli obtained by triangulations are small circles on the celestial
sphere, so their curvature, even across a relatively small BATSE error circle, is
not always negligible, and a simple, four-sided error box cannot be defined. 
This curvature can be seen in Figure 5. 
For this reason, we do not cite the intersection points of
the annulus with the error circle.  However, we give here the formulas for finding
these points for those cases where it may be useful.  Let $\alpha_1, \delta_1, \theta_1$
and $\alpha_2, \delta_2, \theta_2$ be the right ascension, declination, and radii of
the two small circles.  Let $\alpha, \delta$ be the right ascension and declination
of the intersection points.  Then

$\sin\delta=\frac{-b \pm \sqrt{b^2 - 4ac}}{2a}$

where

a=$-\cos^2\delta_1 \cos^2\delta_2 \sin^2(\alpha_1-\alpha_2) + 
2\sin\delta_1 \sin\delta_2 \cos\delta_1 \cos\delta_2 \cos(\alpha_1-\alpha_2) -
\sin^2\delta_2 \cos^2\delta_1 -
\sin^2\delta_1 \cos^2\delta_2$,

b=$-2(\cos\theta_1 \sin\delta_2 + \cos\theta_2 \sin\delta_1) \cos\delta_1 \cos\delta_2 \cos(\alpha_1 - \alpha_2)+
2\cos\theta_2 \sin\delta_2 \cos^2\delta_1 +
2\cos\theta_1 \sin\delta_1 \cos^2\delta_2$,

and

c=$\cos^2\delta_1 \cos^2\delta_2 \sin^2(\alpha_1-\alpha_2) +
2\cos\theta_2 \cos\theta_1 \cos\delta_1 \cos\delta_2 \cos(\alpha_1 - \alpha_2) -
\cos^2\theta_2 \cos^2\delta_1 -
\cos^2\theta_1 \cos^2\delta_2$,

For each of the two values of the declination $\delta$ the corresponding right ascension
$\alpha$ is given by

$\alpha=\alpha_1 + \cos^{-1}\frac{\cos\theta_1 - \sin\delta \sin\delta_1}
{\cos\delta \cos\delta_1}$

Figure 8 shows the BATSE peak fluxes and fluences for 162 of the 218 bursts
with flux and fluence entries in the 3B catalog.  There is a slight tendency for 
some of the smaller
fluence events to have relatively large peak fluxes over a time interval comparable
to that of the \it Ulysses \rm real time data, which helps to explain why these bursts were
detected by \it Ulysses \rm.  To estimate the completeness of this catalog, we have
calculated the ratios of the number of bursts detected by \it Ulysses \rm to the number
detected by BATSE above several
flux and fluence thresholds starting with the weakest events in the BATSE catalog; 
we have used the 25-100 keV fluences in erg cm$^{-2}$ and
the peak 25 - 300 keV fluxes over 256 ms in photons cm$^{-2}$ s$^{-1}$.  Since events
detected by a third spacecraft are not included in the present catalog, the numbers
given represent lower limits to the completeness.  Table 1 summarizes the results.

\clearpage
\begin{deluxetable}{cc}
\tablecaption{Completeness above various thresholds}
\tablehead{
\colhead{BATSE Threshold}&\colhead{Approximate completeness (lower limit)}
}
\startdata

1.8 x 10$^{-9}$ erg cm$^{-2}$ & .25  \nl
10$^{-6}$ erg cm$^{-2}$ & .49 \nl
10$^{-5}$ erg cm$^{-2}$ & .74 \nl
.27 photons cm$^{-2}$ s$^{-1}$ & .25 \nl
1 photon cm$^{-2}$ s$^{-1}$ & .34 \nl
10 photons cm$^{-2}$ s$^{-1}$ & .65 \nl
\enddata
\end{deluxetable}
\clearpage

Finally, we have analyzed these annuli for intersections which might provide evidence
of repeating bursts from a single source.  Although there was a large, but statistically
insignificant number of
two annulus intersections, we found only one case where a three-way
intersection almost occurred (BATSE bursts 451, 2286, and 2619); the data are therefore
consistent with no burst repetition. 

\section{Table of Annuli}

The ten columns in table 1 give:
1) the date of the burst, in ddmmyy format, 
2) the Universal Time of the burst at Earth,
3) the BATSE number for the burst,
4) the BATSE right ascension of the center of the error circle (J2000), in degrees,
5) the BATSE declination of the center of the error circle (J2000), in degrees,
6) the total 1 $\sigma$ statistical BATSE error circle radius, in degrees, (the 
approximate total
1$\sigma$ radius is obtained by adding 1.6$^{\rm o}$ in quadrature, but see
Briggs et al. 1998a,b for an improved error model), 
7) the right ascension of the center of the IPN (BATSE/\it Ulysses \rm)
annulus, epoch J2000, corrected to the heliocentric frame, in degrees,
8) the declination of the center of the IPN (BATSE/\it Ulysses \rm)
annulus, epoch J2000, corrected to the heliocentric frame, in degrees, 
9) the angular radius of the IPN (BATSE/\it Ulysses \rm) annulus, corrected to the heliocentric
frame, in degrees, and
10) the half width of the IPN (BATSE/\it Ulysses \rm) annulus, in degrees; the 3 $\sigma$
confidence annulus is given by R$_{\rm IPN}$ $\pm$ $\delta$ R$_{IPN}$.

This list supercedes previous ones which have been circulated in the past, but
not published.  The changes to it include:
1) the use of final time history and spacecraft ephemeris data; earlier lists
often used preliminary and predict data, with correspondingly larger timing uncertainties,
2) transformation of annuli to heliocentric coordinates; earlier lists often gave
geocentric coordinates, with correspondingly larger uncertainties in the annulus widths,
3) removal of weak events according to the criteria in section 5,
4) reprocessing of data to eliminate known clock errors, and
5) careful examination of bursts whose BATSE time histories were truncated
by data gaps and Earth occultation.   

Figures 9 and 10 compare the BATSE error circles with the IPN annulus/error circle
intersections for the bursts in this catalog.  To generate this plot, it was 
assumed that all annuli pass through the centers of their corresponding BATSE error 
circles.
 
We stress that while the BATSE data are for the bursts listed in the 3B
catalog, the locations and error circle radii are in fact the updated ones
which will appear in the 4B catalog.  They have been taken from the latest online catalog,
and are given here for convenience only. \footnote{
The catalog (http://www.batse.msfc.nasa.gov/data/grb/4bcatalog/)
should be considered to be the ultimate source of the
most up-to-date BATSE data.   Table 1 is available electronically from
http://ssl.berkeley.edu/ipn3/index.html.}  Finally, we add that 150 \it Ulysses \rm/BATSE
bursts were detected over the period between the end of the 3B catalog
and the end of the 4B catalog.  Analysis of these events has been completed and
is being published in conjunction with the 4B catalog. 

\clearpage
\begin{deluxetable}{cccccccccc}
\tablecaption{\it Ulysses \rm/BATSE annuli}
\tablehead{
\colhead{Date}&\colhead{UT}&\colhead{N$_B$}&\colhead{$\alpha_{2000, B}$}&
\colhead{$\delta_{2000, B}$}&\colhead{$\sigma_{stat, B}$}&\colhead{$\alpha_{2000, IPN}$}&
\colhead{$\delta_{2000, IPN}$}&\colhead{R$_{IPN}$}&\colhead{$\delta R_{IPN}$}
}
\startdata
250491 & 00:37:46 & 109 & 91.29 & -22.77 & 1.02 & 115.140 & 23.171 & 51.784 & 0.051 \nl
290491 & 03:11:50 & 121 & 174.75 & 14.33 & 1.36 & 116.426 & 22.919 & 59.157 & 0.136 \nl
020591 & 22:37:23 & 142 & 46.22 & -52.71 & 0.62 & 297.624 & -22.677 & 87.126 & 0.023 \nl
030591 & 07:04:13 & 143 & 87.45 & 38.74 & 0.87 & 117.731 & 22.655 & 30.360 & 0.009 \nl
110591 & 02:11:48 & 179 & 266.48 & 58.30 & 1.66 & 300.193 & -22.137 & 83.666 & 0.018 \nl
230591 & 19:03:25 & 222 & 106.99 & 0.36 & 1.28 & 124.202 & 21.228 & 26.847 & 0.051 \nl
010691 & 19:22:14 & 249 & 310.12 & 32.34 & 0.17 & 307.035 & -20.537 & 52.942 & 0.008 \nl
020691 & 22:55:01 & 257 & 142.76 & 53.99 & 2.06 & 127.387 & 20.448 & 34.526 & 0.163 \nl
260691 & 07:15:13 & 444 & 133.38 & 7.72 & 1.27 & 134.557 & 18.501 & 10.164 & 0.020 \nl
300691 & 07:37:02 & 469 & 304.70 & 35.13 & 0.91 & 315.764 & -18.148 & 52.439 & 0.085 \nl
210791 & 19:30:12 & 563 & 315.52 & -14.29 & 1.31 & 322.022 & -16.198 & 6.769 & 0.436 \nl
090891 & 06:15:00 & 659 & 195.43 & 19.37 & 1.85 & 147.088 & 14.483 & 50.860 & 0.092 \nl
090891 & 19:35:45 & 660 & 281.96 & 64.74 & 2.01 & 327.238 & -14.431 & 85.747 & 0.016 \nl
050991 & 23:48:55 & 761 & 360.00 & -81.05 & 0.82 & 334.075 & -11.942 & 70.787 & 0.017 \nl
070991 & 01:05:23 & 764 & 136.55 & -28.23 & 1.30 & 154.322 & 11.849 & 43.573 & 0.131 \nl
260991 & 18:25:57 & 824 & 346.07 & 21.36 & 2.70 & 338.688 & -10.171 & 32.155 & 0.294 \nl
270991 & 23:26:53 & 829 & 50.16 & -39.65 & 0.34 & 338.937 & -10.074 & 68.870 & 0.036 \nl
300991 & 11:55:30 & 841 & 131.80 & -20.62 & 1.27 & 159.447 & 9.874 & 39.201 & 0.013 \nl
061091 & 09:01:54 & 871 & 33.30 & -45.44 & 3.38 & 340.603 & -9.419 & 50.544 & 0.087 \nl
221091 & 04:13:58 & 914 & 123.91 & 70.34 & 1.03 & 163.395 & 8.312 & 68.583 & 0.017 \nl
261091 & 13:07:01 & 938 & 74.68 & -22.27 & 2.40 & 344.080 & -8.038 & 80.644 & 0.068 \nl
061191 & 03:42:01 & 1008 & 343.03 & -35.16 & 0.83 & 345.570 & -7.445 & 29.223 & 0.023 \nl
231191 & 08:15:33 & 1114 & 22.81 & 30.34 & 1.14 & 347.386 & -6.730 & 49.421 & 0.015 \nl
271191 & 04:22:08 & 1122 & 269.14 & 49.73 & 0.38 & 347.680 & -6.617 & 88.466 & 0.005 \nl
071291 & 09:50:45 & 1150 & 307.79 & 29.30 & 1.59 & 348.248 & -6.408 & 53.876 & 0.059 \nl
091291 & 18:35:58 & 1157 & 259.98 & -45.08 & 0.40 & 348.333 & -6.380 & 82.576 & 0.003 \nl
171291 & 08:22:06 & 1190 & 6.96 & 22.67 & 1.02 & 348.480 & -6.340 & 35.915 & 0.032 \nl
211291 & 22:03:26 & 1200 & 215.82 & -42.72 & 1.98 & 168.475 & 6.356 & 61.764 & 0.111 \nl
241291 & 11:02:58 & 1212 & 45.43 & -10.79 & 4.68 & 348.444 & -6.377 & 57.290 & 0.117 \nl
271291 & 01:05:53 & 1235 & 350.44 & -68.25 & 0.67 & 348.389 & -6.408 & 61.852 & 0.064 \nl
100192 & 09:17:58 & 1288 & 58.08 & -20.80 & 0.42 & 347.700 & -6.748 & 69.190 & 0.018 \nl
210292 & 06:14:32 & 1425 & 184.14 & 48.48 & 0.65 & 161.368 & 8.446 & 42.142 & 0.008 \nl
240292 & 02:43:39 & 1432 & 169.65 & -66.78 & 2.73 & 160.726 & 8.515 & 75.175 & 0.539 \nl
270292 & 06:00:51 & 1443 & 68.36 & 66.84 & 1.33 & 160.016 & 8.591 & 83.353 & 0.005 \nl
070392 & 00:18:04 & 1467 & 355.57 & -45.15 & 1.16 & 338.049 & -8.795 & 39.316 & 0.066 \nl
150392 & 04:19:27 & 1484 & 320.58 & -20.89 & 0.46 & 336.280 & -8.960 & 16.873 & 0.013 \nl
070492 & 12:03:40 & 1543 & 341.44 & -10.68 & 2.04 & 332.036 & -9.187 & 9.927 & 0.437 \nl
080492 & 12:05:56 & 1545 & 138.96 & 40.83 & 0.58 & 151.890 & 9.190 & 34.380 & 0.022 \nl
290492 & 18:59:32 & 1571 & 179.08 & -14.33 & 2.85 & 149.602 & 8.963 & 39.318 & 0.091 \nl
020592 & 06:02:59 & 1577 & 33.05 & -17.73 & 1.04 & 329.440 & -8.908 & 60.781 & 0.026 \nl
130592 & 16:52:39 & 1606 & 211.16 & -44.77 & 0.65 & 148.956 & 8.591 & 74.250 & 0.005 \nl
300592 & 22:59:55 & 1630 & 162.68 & 18.13 & 1.67 & 148.993 & 7.901 & 17.618 & 0.070 \nl
170692 & 05:26:51 & 1652 & 37.66 & 77.51 & 0.51 & 149.807 & 6.977 & 89.064 & 0.012 \nl
220692 & 07:05:05 & 1663 & 162.10 & 47.17 & 0.31 & 150.169 & 6.667 & 41.336 & 0.005 \nl
220692 & 12:54:21 & 1664 & 346.81 & 10.01 & 0.55 & 330.187 & -6.649 & 18.892 & 0.018 \nl
270692 & 13:02:36 & 1676 & 166.02 & -2.58 & 0.65 & 150.593 & 6.327 & 23.504 & 0.017 \nl
010792 & 19:38:19 & 1683 & 314.64 & 33.44 & 0.63 & 330.973 & -6.033 & 42.560 & 0.017 \nl
140792 & 20:23:26 & 1700 & 140.36 & -44.64 & 3.28 & 152.298 & 5.079 & 50.262 & 0.099 \nl
180792 & 14:40:36 & 1708 & 22.44 & -3.98 & 1.14 & 332.718 & -4.782 & 49.196 & 0.032 \nl
180792 & 21:32:43 & 1709 & 296.39 & -55.98 & 0.48 & 332.751 & -4.760 & 58.948 & 0.004 \nl
210792 & 18:17:48 & 1717 & 37.55 & 33.49 & 0.73 & 333.083 & -4.530 & 72.485 & 0.033 \nl
230792 & 01:00:34 & 1721 & 85.79 & 7.78 & 0.48 & 153.234 & 4.427 & 68.971 & 0.008 \nl
010892 & 01:16:40 & 1733 & 157.86 & 68.90 & 1.12 & 154.338 & 3.667 & 64.462 & 0.021 \nl
130892 & 03:33:48 & 1807 & 206.22 & -28.34 & 1.75 & 155.907 & 2.581 & 63.921 & 0.051 \nl
140892 & 06:09:53 & 1815 & 258.68 & -44.63 & 0.79 & 336.053 & -2.478 & 78.143 & 0.006 \nl
240892 & 10:53:03 & 1872 & 228.59 & -51.79 & 0.33 & 157.427 & 1.505 & 76.788 & 0.003 \nl
300892 & 01:45:18 & 1883 & 258.53 & -73.99 & 0.54 & 338.192 & -0.949 & 86.267 & 0.010 \nl
020992 & 00:29:05 & 1886 & 279.15 & -20.70 & 2.29 & 338.594 & -0.651 & 61.607 & 0.015 \nl
030992 & 00:49:17 & 1888 & 61.97 & -68.37 & 0.75 & 338.732 & -0.549 & 88.165 & 0.010 \nl
250992 & 21:45:16 & 1956 & 160.26 & 67.38 & 1.34 & 161.773 & -1.864 & 68.233 & 0.020 \nl
031092 & 06:35:40 & 1974 & 150.32 & -14.71 & 0.55 & 162.694 & -2.674 & 19.765 & 0.021 \nl
091092 & 06:41:11 & 1983 & 120.41 & -29.10 & 0.39 & 163.411 & -3.345 & 46.240 & 0.003 \nl
151092 & 01:36:32 & 1989 & 123.04 & 41.69 & 0.95 & 164.066 & -3.998 & 57.677 & 0.040 \nl
221092 & 15:20:59 & 1997 & 253.78 & -12.42 & 0.41 & 164.862 & -4.861 & 88.377 & 0.002 \nl
291092 & 12:37:56 & 2018 & 34.14 & -1.57 & 1.67 & 345.514 & 5.651 & 50.169 & 0.045 \nl
301092 & 01:16:31 & 2020 & 86.16 & 6.86 & 1.75 & 165.561 & -5.710 & 79.293 & 0.042 \nl
181192 & 22:11:44 & 2061 & 350.45 & 50.32 & 1.07 & 346.930 & 7.990 & 41.232 & 0.128 \nl
231192 & 06:18:31 & 2067 & 334.24 & -53.25 & 0.72 & 347.111 & 8.481 & 64.897 & 0.008 \nl
031292 & 21:44:22 & 2074 & 233.62 & 21.03 & 0.74 & 167.337 & -9.664 & 68.339 & 0.119 \nl
061292 & 18:24:19 & 2080 & 176.48 & 45.72 & 0.77 & 167.341 & -9.975 & 51.667 & 0.029 \nl
071292 & 16:00:48 & 2083 & 306.33 & -42.93 & 0.44 & 347.336 & 10.074 & 62.709 & 0.003 \nl
091292 & 02:40:56 & 2089 & 172.98 & 61.45 & 0.51 & 167.325 & -10.228 & 68.392 & 0.048 \nl
091292 & 11:35:38 & 2090 & 332.97 & -52.02 & 0.68 & 347.320 & 10.269 & 63.995 & 0.009 \nl
181292 & 02:30:02 & 2102 & 355.62 & -55.31 & 1.76 & 347.091 & 11.168 & 66.903 & 0.076 \nl
271292 & 12:44:08 & 2106 & 206.33 & -25.65 & 1.04 & 166.535 & -12.089 & 39.698 & 0.113 \nl
301292 & 08:57:51 & 2110 & 19.52 & 17.21 & 0.33 & 346.301 & 12.353 & 32.939 & 0.112 \nl
301292 & 14:05:46 & 2111 & 150.54 & 28.63 & 1.71 & 166.282 & -12.370 & 45.110 & 0.086 \nl
060193 & 15:37:39 & 2121 & 6.08 & 2.00 & 0.53 & 345.556 & 12.984 & 21.048 & 0.021 \nl
060193 & 19:57:15 & 2122 & 124.79 & -32.62 & 0.68 & 165.536 & -12.996 & 42.302 & 0.149 \nl
080193 & 02:28:55 & 2123 & 94.62 & -30.44 & 0.99 & 165.383 & -13.100 & 65.116 & 0.069 \nl
120193 & 03:47:31 & 2127 & 226.32 & 27.74 & 1.22 & 164.853 & -13.415 & 72.911 & 0.021 \nl
120193 & 15:18:02 & 2128 & 210.10 & 51.92 & 1.02 & 164.786 & -13.451 & 76.849 & 0.066 \nl
160193 & 02:47:02 & 2136 & 246.99 & 15.91 & 1.02 & 164.275 & -13.700 & 88.748 & 0.026 \nl
270193 & 16:14:25 & 2149 & 310.95 & -39.63 & 1.16 & 342.242 & 14.382 & 60.557 & 0.155 \nl
310193 & 18:57:12 & 2151 & 182.04 & -8.24 & 2.16 & 161.406 & -14.563 & 22.249 & 0.009 \nl
010293 & 16:41:55 & 2156 & 333.53 & 41.15 & 1.26 & 341.214 & 14.601 & 27.869 & 0.023 \nl
050293 & 10:10:31 & 2166 & 278.00 & -45.74 & 5.98 & 340.403 & 14.729 & 83.936 & 0.277 \nl
170293 & 14:44:38 & 2197 & 117.78 & 0.98 & 2.27 & 157.517 & -14.945 & 41.809 & 0.234 \nl
020393 & 21:33:17 & 2213 & 90.33 & 5.43 & 0.62 & 154.166 & -14.821 & 66.384 & 0.016 \nl
090393 & 03:07:50 & 2228 & 325.41 & 51.73 & 0.72 & 332.605 & 14.648 & 36.594 & 0.012 \nl
100393 & 07:19:20 & 2232 & 284.35 & -52.39 & 0.89 & 332.316 & 14.607 & 79.362 & 0.007 \nl
180393 & 12:29:53 & 2255 & 150.48 & -18.85 & 1.09 & 150.361 & -14.267 & 5.323 & 0.320 \nl
290393 & 03:09:25 & 2273 & 179.25 & -8.65 & 0.94 & 148.090 & -13.723 & 35.582 & 0.010 \nl
310393 & 03:11:16 & 2276 & 43.00 & 47.27 & 2.03 & 327.701 & 13.613 & 77.070 & 0.078 \nl
050493 & 08:54:49 & 2286 & 198.18 & -8.80 & 0.55 & 146.754 & -13.315 & 52.136 & 0.005 \nl
060493 & 07:16:25 & 2289 & 299.81 & 29.01 & 0.51 & 326.595 & 13.263 & 31.383 & 0.011 \nl
090493 & 04:50:10 & 2295 & 295.83 & 43.18 & 2.04 & 326.125 & 13.098 & 37.466 & 0.366 \nl
090493 & 21:25:40 & 2297 & 176.91 & -39.75 & 1.25 & 146.019 & -13.056 & 39.155 & 0.088 \nl
140493 & 02:27:15 & 2303 & 311.42 & 9.92 & 0.91 & 325.406 & 12.819 & 14.314 & 0.061 \nl
180493 & 14:38:18 & 2307 & 311.65 & 21.37 & 1.27 & 324.831 & 12.570 & 14.596 & 0.415 \nl
240493 & 19:19:00 & 2315 & 54.22 & 41.60 & 1.67 & 324.179 & 12.245 & 86.680 & 0.134 \nl
250493 & 10:17:29 & 2316 & 18.00 & -35.19 & 0.31 & 324.122 & 12.214 & 70.285 & 0.034 \nl
260493 & 12:40:32 & 2319 & 67.82 & 10.13 & 1.10 & 144.025 & -12.159 & 75.515 & 0.028 \nl
280493 & 01:07:12 & 2320 & 129.85 & -55.45 & 1.68 & 143.900 & -12.085 & 46.202 & 0.005 \nl
300493 & 15:01:18 & 2321 & 106.33 & 20.97 & 0.36 & 143.709 & -11.967 & 50.803 & 0.009 \nl
020593 & 13:52:52 & 2323 & 254.72 & -46.53 & 0.78 & 323.580 & 11.882 & 87.103 & 0.039 \nl
060593 & 14:52:51 & 2329 & 64.45 & -5.97 & 0.24 & 143.364 & -11.717 & 77.827 & 0.004 \nl
180593 & 03:20:25 & 2346 & 349.17 & -37.51 & 1.04 & 323.086 & 11.354 & 53.387 & 0.019 \nl
230593 & 19:40:53 & 2350 & 14.80 & 4.21 & 2.64 & 323.121 & 11.238 & 44.153 & 0.124 \nl
310593 & 15:53:38 & 2362 & 340.04 & 13.25 & 2.16 & 323.341 & 11.152 & 15.895 & 0.044 \nl
090693 & 10:07:24 & 2383 & 208.73 & -16.38 & 1.62 & 143.799 & -11.158 & 64.377 & 0.132 \nl
140693 & 03:40:30 & 2393 & 168.14 & 23.95 & 1.15 & 144.132 & -11.210 & 42.800 & 0.008 \nl
050793 & 12:39:08 & 2429 & 283.08 & -39.24 & 0.99 & 326.244 & 11.861 & 65.410 & 0.124 \nl
060793 & 08:09:10 & 2432 & 210.09 & -1.36 & 1.31 & 146.342 & -11.898 & 63.498 & 0.045 \nl
080793 & 05:12:38 & 2433 & 86.55 & -69.02 & 6.25 & 146.572 & -11.989 & 62.867 & 0.076 \nl
090793 & 06:14:22 & 2436 & 194.41 & -46.73 & 0.67 & 146.700 & -12.042 & 54.525 & 0.028 \nl
140793 & 16:12:53 & 2446 & 257.70 & 23.70 & 0.97 & 327.399 & 12.344 & 68.720 & 0.048 \nl
210793 & 01:17:22 & 2452 & 78.66 & -33.15 & 1.52 & 148.275 & -12.751 & 66.262 & 0.085 \nl
210793 & 12:45:35 & 2453 & 298.58 & -40.79 & 1.79 & 328.342 & 12.785 & 61.079 & 0.101 \nl
270793 & 08:08:29 & 2466 & 159.99 & -23.89 & 1.28 & 149.185 & -13.209 & 16.546 & 0.136 \nl
310793 & 03:16:31 & 2475 & 116.53 & 45.49 & 2.15 & 149.754 & -13.514 & 66.345 & 0.036 \nl
050893 & 12:39:28 & 2482 & 334.55 & -15.60 & 0.83 & 330.584 & 13.983 & 30.305 & 0.208 \nl
090893 & 05:11:04 & 2486 & 37.97 & -74.63 & 1.79 & 151.166 & -14.323 & 80.857 & 0.088 \nl
220893 & 14:44:51 & 2500 & 213.09 & -0.68 & 2.28 & 153.348 & -15.724 & 65.872 & 0.232 \nl
310893 & 09:16:05 & 2507 & 223.29 & -3.63 & 0.47 & 154.815 & -16.774 & 67.536 & 0.007 \nl
030993 & 00:19:29 & 2511 & 88.67 & -10.42 & 4.20 & 155.258 & -17.109 & 64.672 & 0.109 \nl
050993 & 03:26:30 & 2514 & 310.95 & -1.22 & 0.63 & 335.615 & 17.389 & 28.629 & 0.006 \nl
100993 & 12:12:08 & 2522 & 167.59 & -66.69 & 0.54 & 156.521 & -18.114 & 48.810 & 0.044 \nl
140993 & 02:49:11 & 2530 & 34.26 & 34.29 & 1.53 & 337.128 & 18.628 & 50.198 & 0.173 \nl
160993 & 20:19:23 & 2533 & 280.71 & 65.39 & 0.22 & 337.586 & 19.026 & 57.392 & 0.013 \nl
220993 & 06:24:46 & 2537 & 271.20 & 55.89 & 0.45 & 338.488 & 19.847 & 67.999 & 0.003 \nl
270993 & 04:18:07 & 2542 & 108.61 & -9.52 & 1.62 & 159.296 & -20.624 & 52.084 & 0.115 \nl
061093 & 21:31:50 & 2566 & 66.31 & 65.28 & 1.70 & 340.850 & 22.262 & 70.461 & 0.049 \nl
081093 & 07:18:14 & 2570 & 42.98 & 21.37 & 0.85 & 341.069 & 22.510 & 58.202 & 0.082 \nl
081093 & 11:09:10 & 2571 & 165.40 & 33.93 & 0.86 & 161.094 & -22.537 & 60.104 & 0.027 \nl
141093 & 17:01:01 & 2586 & 272.50 & 9.34 & 0.50 & 342.042 & 23.672 & 67.956 & 0.010 \nl
171093 & 01:17:49 & 2590 & 181.44 & -83.11 & 0.65 & 162.389 & -24.110 & 58.979 & 0.003 \nl
191093 & 18:22:34 & 2593 & 242.92 & -18.70 & 1.66 & 162.779 & -24.628 & 73.476 & 0.025 \nl
241093 & 13:29:09 & 2603 & 137.63 & -8.49 & 0.53 & 163.447 & -25.567 & 30.574 & 0.083 \nl
261093 & 11:35:57 & 2606 & 51.36 & -10.56 & 0.48 & 343.704 & 25.955 & 75.482 & 0.087 \nl
301093 & 08:08:42 & 2609 & 326.41 & 58.24 & 0.55 & 344.202 & 26.745 & 31.863 & 0.058 \nl
311093 & 04:06:35 & 2611 & 325.10 & 62.73 & 0.32 & 344.306 & 26.918 & 37.073 & 0.005 \nl
031193 & 16:25:42 & 2617 & 87.88 & 65.05 & 0.31 & 344.732 & 27.657 & 72.278 & 0.008 \nl
061193 & 20:32:04 & 2619 & 196.75 & -3.13 & 1.11 & 165.094 & -28.337 & 40.406 & 0.232 \nl
121193 & 18:45:53 & 2628 & 311.51 & 55.95 & 0.53 & 345.711 & 29.654 & 34.291 & 0.010 \nl
261193 & 04:46:46 & 2660 & 157.28 & 70.08 & 3.02 & 346.750 & 32.795 & 80.134 & 0.109 \nl
261193 & 19:31:39 & 2661 & 3.29 & -17.40 & 0.22 & 346.785 & 32.945 & 50.498 & 0.004 \nl
041293 & 09:48:10 & 2676 & 250.15 & 38.23 & 0.18 & 347.083 & 34.824 & 76.909 & 0.004 \nl
051293 & 14:59:51 & 2679 & 15.56 & 66.49 & 2.22 & 347.108 & 35.133 & 38.364 & 0.018 \nl
081293 & 04:37:19 & 2682 & 198.73 & -29.93 & 0.56 & 167.142 & -35.783 & 34.384 & 0.050 \nl
081293 & 09:34:40 & 2683 & 234.11 & -40.29 & 3.06 & 167.143 & -35.836 & 51.960 & 0.143 \nl
211293 & 02:09:50 & 2700 & 91.02 & -43.17 & 0.50 & 166.799 & -39.143 & 56.073 & 0.036 \nl
221293 & 21:21:33 & 2703 & 192.52 & 28.65 & 0.66 & 166.675 & -39.621 & 73.261 & 0.251 \nl
251293 & 23:20:12 & 2709 & 4.85 & 9.87 & 0.79 & 346.415 & 40.440 & 35.141 & 0.212 \nl
261293 & 20:11:09 & 2711 & 205.95 & 21.63 & 2.82 & 166.329 & -40.669 & 74.178 & 0.127 \nl
010194 & 22:42:00 & 2729 & 189.27 & 62.84 & 3.47 & 345.578 & 42.289 & 78.226 & 0.064 \nl
020194 & 02:45:52 & 2732 & 10.90 & 45.11 & 1.27 & 345.554 & 42.335 & 18.951 & 0.195 \nl
030194 & 22:12:56 & 2736 & 267.90 & 7.14 & 0.65 & 345.274 & 42.810 & 76.721 & 0.016 \nl
130194 & 16:47:54 & 2756 & 209.39 & -23.16 & 9.65 & 163.289 & -45.332 & 42.651 & 0.049 \nl
190194 & 15:44:40 & 2773 & 131.41 & 39.29 & 1.39 & 161.654 & -46.793 & 89.606 & 0.051 \nl
280194 & 16:50:58 & 2790 & 224.02 & -15.22 & 0.96 & 158.510 & -48.816 & 64.647 & 0.059 \nl
290194 & 10:43:17 & 2793 & 127.46 & 28.59 & 0.82 & 158.217 & -48.971 & 81.612 & 0.072 \nl
010294 & 07:40:40 & 2794 & 268.61 & 22.37 & 1.67 & 337.027 & 49.542 & 59.277 & 0.125 \nl
030294 & 15:46:56 & 2797 & 184.16 & -40.95 & 0.58 & 156.002 & -49.973 & 20.512 & 0.052 \nl
060294 & 00:08:37 & 2798 & 144.20 & -59.96 & 0.15 & 154.922 & -50.380 & 13.687 & 0.039 \nl
100294 & 19:13:16 & 2812 & 152.29 & 82.13 & 0.66 & 332.566 & 51.121 & 48.298 & 0.007 \nl
170294 & 23:02:42 & 2831 & 29.07 & 4.55 & 0.69 & 328.739 & 51.947 & 69.334 & 0.005 \nl
180294 & 19:32:28 & 2833 & 349.71 & -19.69 & 0.35 & 328.264 & 52.023 & 73.939 & 0.010 \nl
280294 & 11:29:00 & 2852 & 127.98 & -12.36 & 0.44 & 142.713 & -52.495 & 42.173 & 0.012 \nl
010394 & 20:10:37 & 2855 & 103.51 & 64.35 & 0.35 & 321.919 & 52.509 & 59.400 & 0.029 \nl
020394 & 05:08:31 & 2856 & 12.49 & -23.97 & 0.21 & 321.700 & 52.508 & 88.572 & 0.005 \nl
120394 & 11:28:22 & 2877 & 221.50 & -56.17 & 0.61 & 135.895 & -52.124 & 48.964 & 0.052 \nl
190394 & 23:57:20 & 2889 & 68.74 & -12.60 & 0.39 & 132.014 & -51.403 & 62.552 & 0.090 \nl
210394 & 22:05:07 & 2890 & 160.33 & 3.10 & 0.85 & 131.093 & -51.168 & 58.918 & 0.233 \nl
230394 & 22:04:38 & 2891 & 213.89 & 8.37 & 0.85 & 310.173 & 50.907 & 85.032 & 0.016 \nl
280394 & 12:05:34 & 2894 & 97.18 & -32.35 & 0.58 & 128.216 & -50.236 & 26.692 & 0.028 \nl
290394 & 05:49:26 & 2895 & 237.51 & -68.89 & 0.50 & 127.917 & -50.121 & 51.112 & 0.006 \nl
290394 & 18:15:37 & 2897 & 204.91 & -10.33 & 1.00 & 127.712 & -50.040 & 69.434 & 0.026 \nl
300394 & 20:44:39 & 2899 & 193.80 & -33.63 & 1.72 & 127.286 & -49.861 & 48.714 & 0.320 \nl
060494 & 01:10:54 & 2913 & 225.44 & -39.10 & 1.24 & 125.133 & -48.806 & 69.435 & 0.043 \nl
100494 & 15:45:02 & 2919 & 254.89 & 30.80 & 0.93 & 303.788 & 47.974 & 42.261 & 0.047 \nl
120494 & 01:40:31 & 2922 & 217.17 & -38.35 & 1.12 & 123.422 & -47.710 & 58.846 & 0.303 \nl
140494 & 16:46:25 & 2929 & 181.26 & -26.15 & 0.50 & 122.794 & -47.222 & 49.154 & 0.026 \nl
190494 & 19:10:58 & 2940 & 359.86 & -47.95 & 0.58 & 121.768 & -46.270 & 73.314 & 0.062 \nl
290494 & 00:43:53 & 2953 & 35.10 & -57.52 & 0.32 & 120.524 & -44.594 & 51.727 & 0.006 \nl
030594 & 05:07:44 & 2958 & 161.00 & 10.06 & 0.60 & 120.195 & -43.873 & 64.039 & 0.057 \nl
120594 & 17:37:50 & 2974 & 232.97 & 55.88 & 2.65 & 299.935 & 42.376 & 44.482 & 0.007 \nl
200594 & 00:21:38 & 2984 & 323.50 & 6.93 & 0.88 & 300.146 & 41.389 & 40.428 & 0.019 \nl
260594 & 10:55:26 & 2993 & 80.41 & -30.39 & 0.96 & 120.590 & -40.650 & 33.571 & 0.347 \nl
260594 & 20:20:05 & 2994 & 131.85 & 34.20 & 1.69 & 120.622 & -40.611 & 76.690 & 0.012 \nl
290594 & 03:17:02 & 2998 & 163.64 & -25.33 & 3.45 & 120.834 & -40.382 & 35.002 & 0.334 \nl
290594 & 16:20:36 & 3002 & 66.19 & -23.75 & 3.30 & 120.889 & -40.330 & 47.443 & 0.021 \nl
290594 & 21:18:56 & 3003 & 214.84 & 58.93 & 0.46 & 300.908 & 40.314 & 56.125 & 0.083 \nl
090694 & 16:29:06 & 3024 & 284.14 & 23.34 & 1.01 & 302.269 & 39.513 & 18.866 & 0.118 \nl
190694 & 21:31:20 & 3035 & 299.05 & -29.89 & 0.34 & 303.996 & 39.141 & 69.575 & 0.026 \nl
220694 & 13:18:16 & 3039 & 299.58 & -31.24 & 0.82 & 304.505 & 39.108 & 72.495 & 0.012 \nl
230694 & 18:46:23 & 3042 & 108.37 & 75.78 & 0.46 & 304.747 & 39.101 & 65.666 & 0.013 \nl
010794 & 21:46:38 & 3055 & 145.18 & -6.43 & 0.81 & 126.479 & -39.199 & 37.139 & 0.157 \nl
020794 & 09:28:11 & 3056 & 29.73 & -23.53 & 0.57 & 126.586 & -39.214 & 78.876 & 0.116 \nl
030794 & 04:40:46 & 3057 & 131.50 & 27.39 & 1.25 & 126.771 & -39.239 & 67.508 & 0.005 \nl
040794 & 23:32:07 & 3060 & 212.49 & 47.25 & 0.64 & 307.181 & 39.305 & 67.688 & 0.082 \nl
060794 & 19:24:33 & 3062 & 265.11 & 26.40 & 1.86 & 307.614 & 39.384 & 36.694 & 0.287 \nl
080794 & 20:42:06 & 3067 & 301.58 & 24.66 & 1.02 & 308.112 & 39.490 & 17.427 & 0.022 \nl
140794 & 12:54:21 & 3075 & 333.58 & -42.10 & 1.59 & 309.555 & 39.862 & 85.495 & 0.067 \nl
170794 & 03:24:29 & 3087 & 109.80 & 12.93 & 0.71 & 130.246 & -40.076 & 54.328 & 0.006 \nl
280794 & 02:55:32 & 3099 & 42.57 & -49.08 & 0.76 & 133.354 & -41.280 & 60.269 & 0.045 \nl
280794 & 14:06:40 & 3100 & 291.22 & 4.45 & 1.29 & 313.491 & 41.344 & 44.189 & 0.181 \nl
280794 & 23:58:54 & 3101 & 85.49 & -39.87 & 1.44 & 133.620 & -41.397 & 36.263 & 0.066 \nl
060894 & 04:38:48 & 3108 & 276.21 & -31.33 & 10.69 & 316.177 & 42.651 & 86.962 & 0.079 \nl
060894 & 09:32:56 & 3109 & 242.38 & 13.65 & 1.22 & 316.242 & 42.686 & 69.105 & 0.126 \nl
100894 & 02:22:42 & 3115 & 212.01 & -17.47 & 0.53 & 137.468 & -43.347 & 67.696 & 0.009 \nl
120894 & 00:27:42 & 3119 & 340.49 & -46.12 & 2.23 & 138.120 & -43.715 & 86.692 & 0.263 \nl
170894 & 08:40:15 & 3128 & 280.84 & 3.47 & 0.40 & 319.994 & 44.823 & 53.907 & 0.010 \nl
210894 & 21:51:27 & 3131 & 294.69 & 43.48 & 1.77 & 321.670 & 45.864 & 16.142 & 0.191 \nl
260894 & 20:55:20 & 3138 & 93.59 & -34.63 & 0.45 & 143.593 & -47.103 & 37.845 & 0.013 \nl
300894 & 08:59:56 & 3143 & 193.66 & 25.66 & 1.58 & 145.011 & -48.048 & 89.313 & 0.042 \nl
100994 & 19:31:24 & 3163 & 332.01 & -34.11 & 0.94 & 330.099 & 51.525 & 87.032 & 0.065 \nl
150994 & 06:52:19 & 3168 & 219.42 & 47.70 & 4.53 & 332.318 & 53.052 & 62.836 & 0.393 \nl\nl

\enddata
\end{deluxetable}
\clearpage

\section{Acknowledgments}

Support for the \it Ulysses \rm GRB experiment is provided by JPL Contract 958056.  Joint
analysis of \it Ulysses \rm and BATSE data is supported by NASA Grant NAG 5-1560.  CK acknowledges
support from NASA Grant NAG5-2560.

\clearpage

\begin{figure}
\plotone{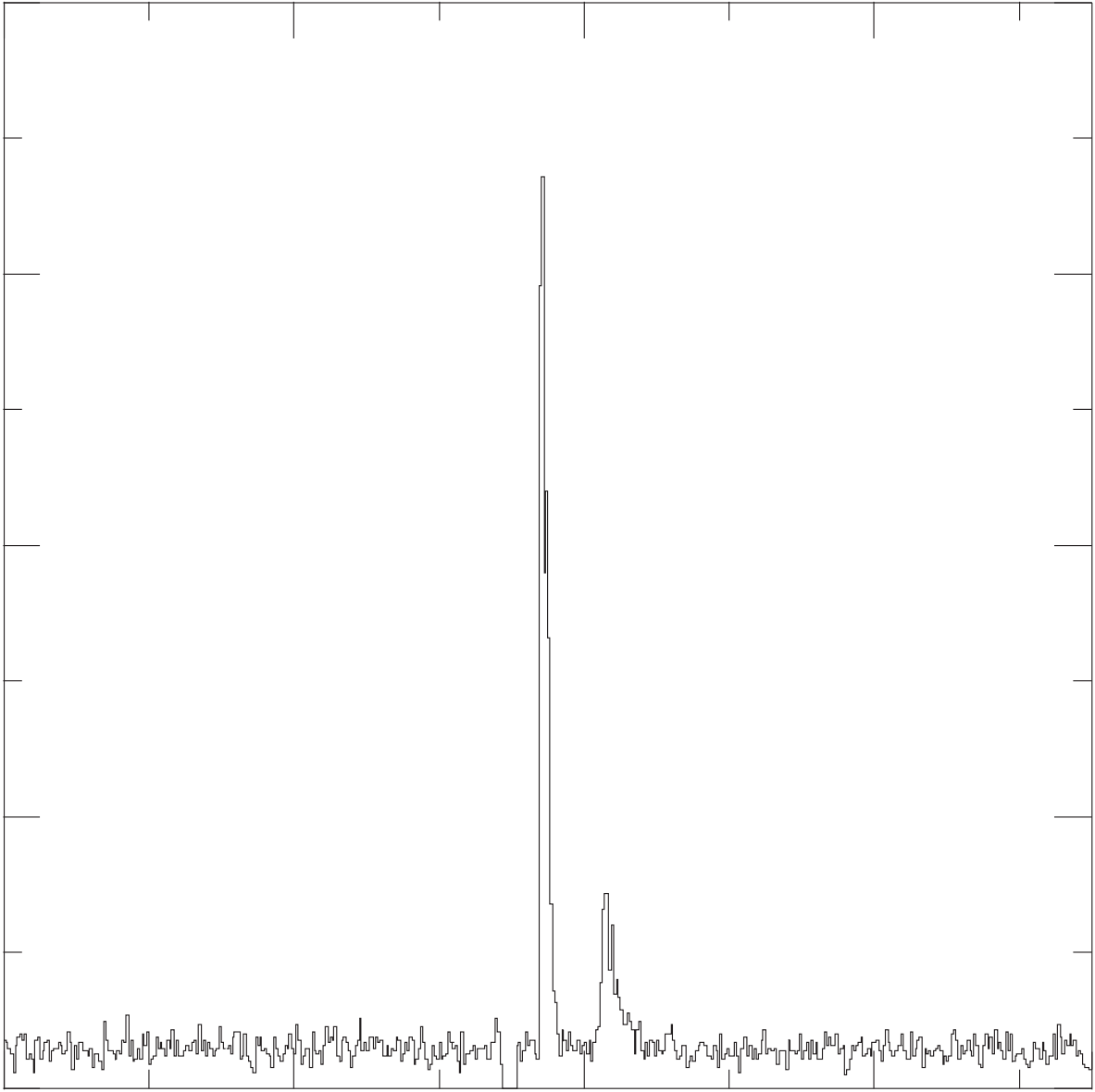}
\caption{\it Ulysses \rm triggered data.  The response to 
BATSE burst 2151 on 1993 January 31 is shown.  The time resolution is 0.03125 s.
\label{Fig. 1}}
\end{figure}

\begin{figure}
\plotone{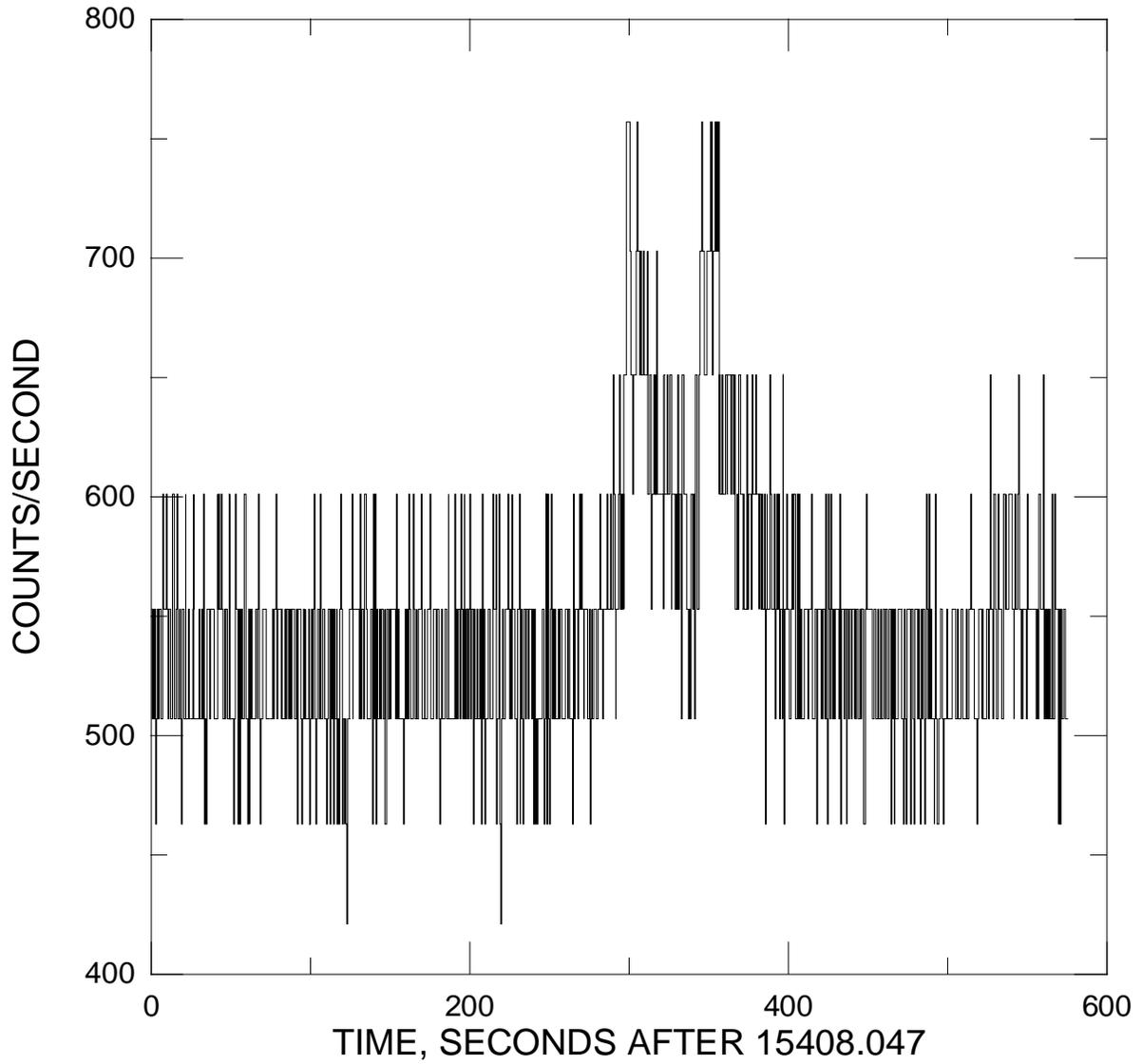}
\caption{\it Ulysses \rm GRB 0.5 s resolution real time data for the crossing window
for BATSE burst 1008 on 1991 November 6.  BATSE observed only the first
peak of this event before entering the South Atlantic Anomaly.  \it Ulysses \rm GRB
triggered on the second peak.  The ``quantized'' appearance of the data is
due to compression.  \label{Fig. 2}}
\end{figure}

\begin{figure}
\plotone{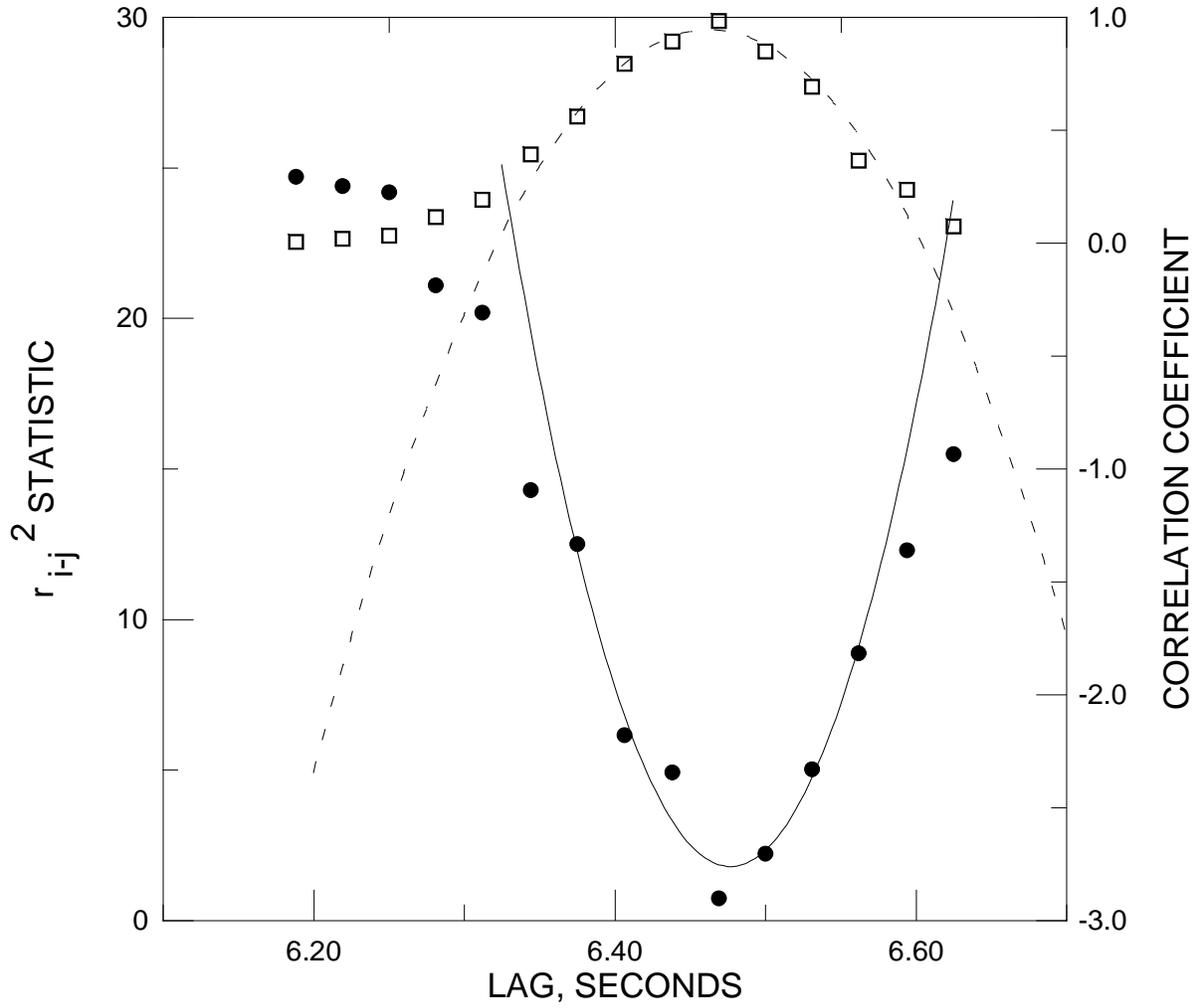}
\caption{r$_{ij}^2$ (dots) and correlation coefficient $\rho_{ij}$
(squares) as a function of lag for
the burst of Figure 1, and their parabolic fits.  The 3$\sigma$ confidence interval
corresponds to $\approx$ 6 ms.  \label{Fig. 3}}
\end{figure}

\begin{figure}
\plotone{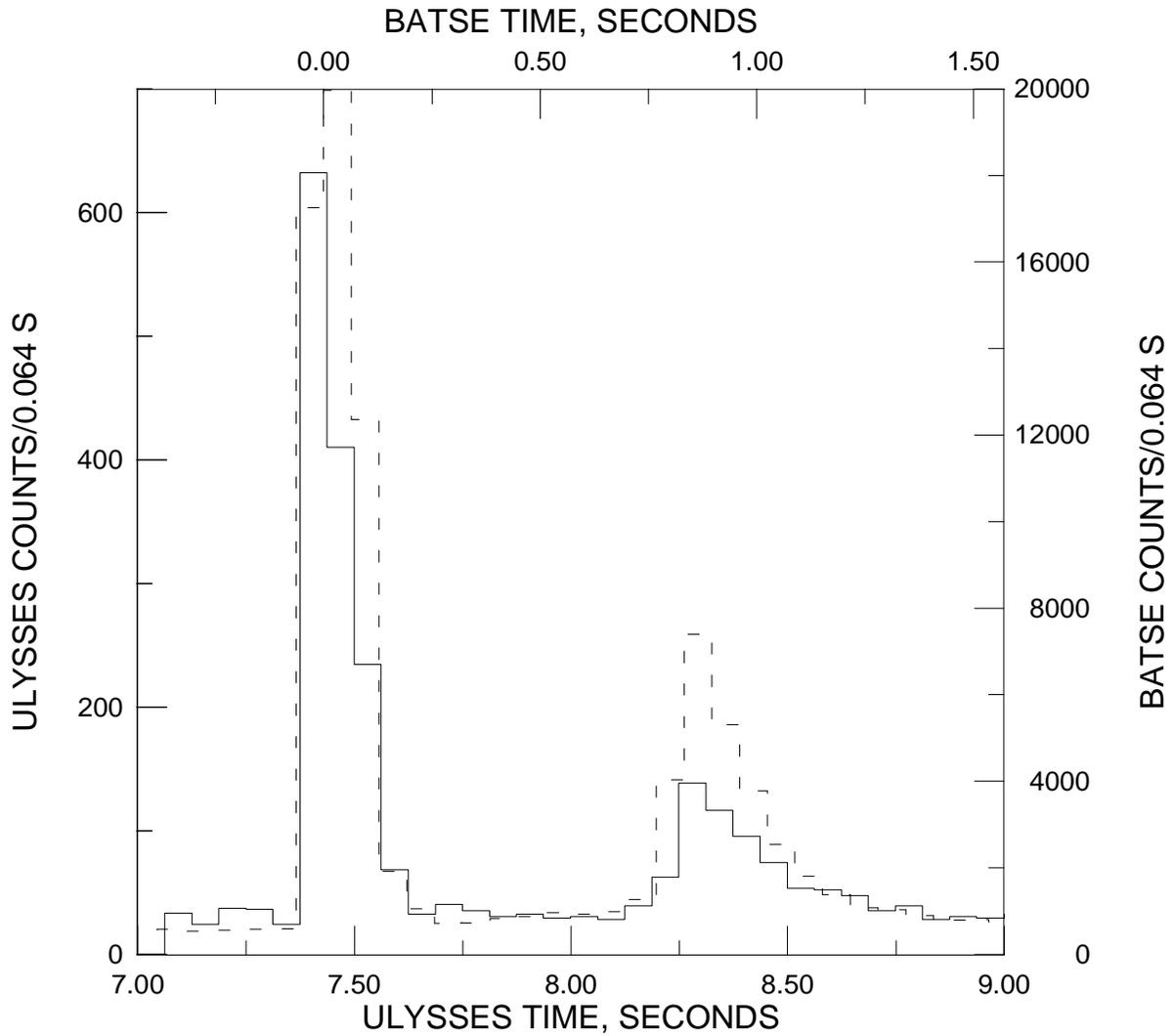}
\caption{\it Ulysses \rm (solid line) and BATSE (dashed line) time histories aligned
for the best fit, for the burst of Figure 1. The BATSE data are summed over 4 LAD's and
are not corrected for deadtime.  \label{Fig. 4}}
\end{figure}

\begin{figure}
\plotone{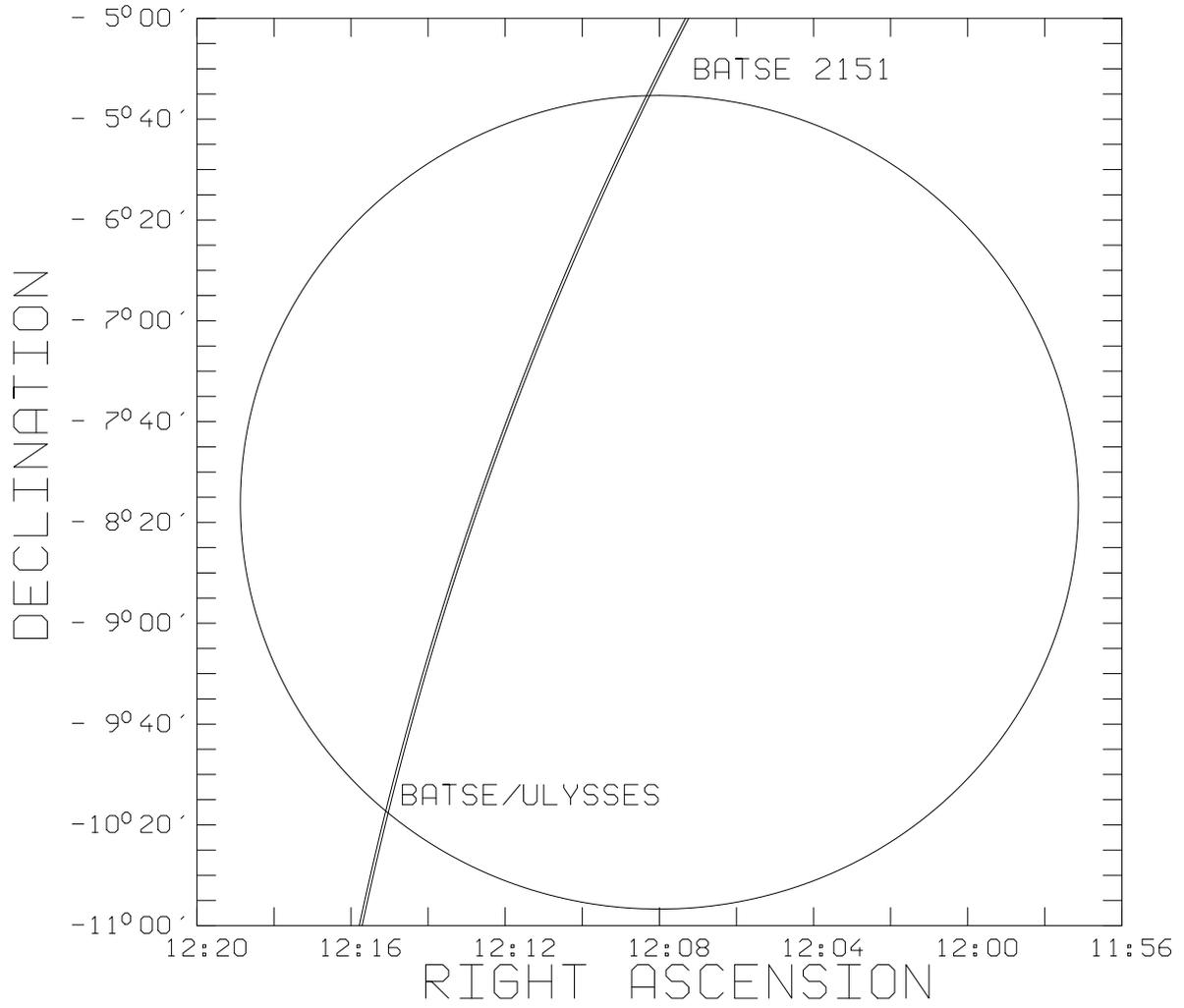}
\caption{The \it Ulysses \rm/BATSE triangulation annulus for the burst of Figure 1,
and the BATSE error circle.  \label{Fig. 5}}
\end{figure}

\begin{figure}
\plotone{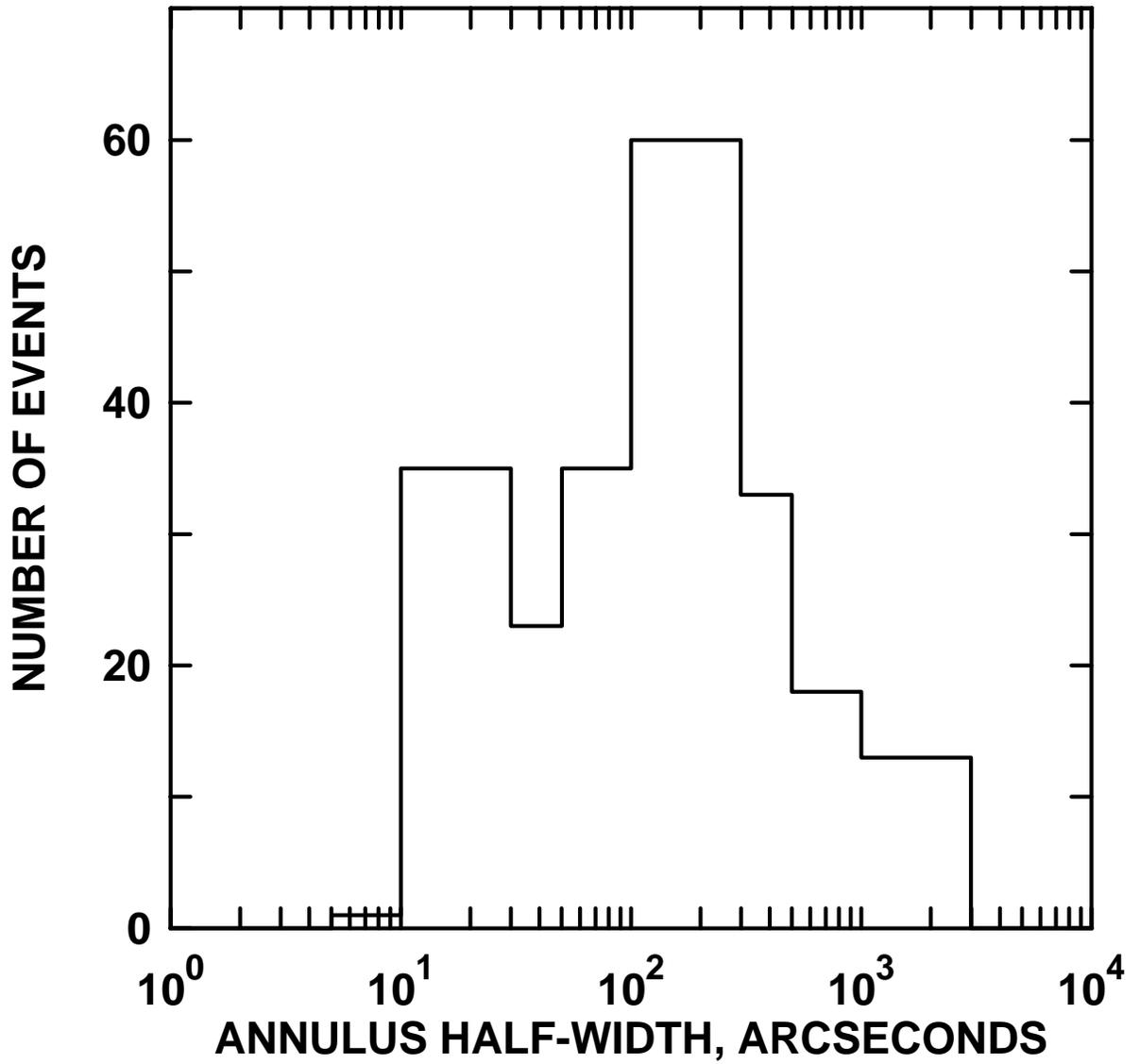}
\caption{Distribution of 218 \it Ulysses \rm/BATSE annulus half-widths (i.e.,
$\delta R_{IPN}$ in Table 1). \label{Fig. 6}}
\end{figure}

\begin{figure}
\plotone{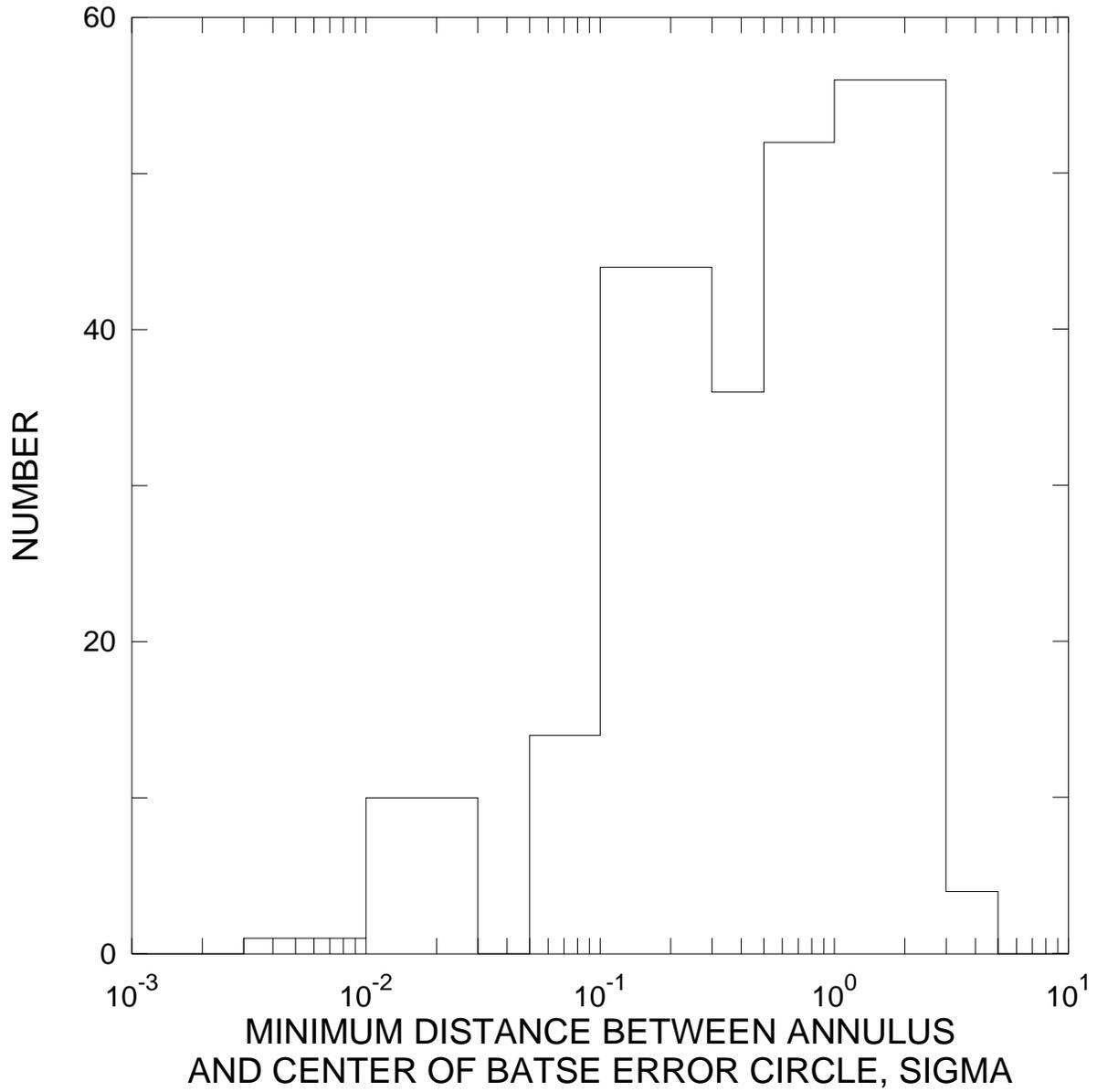}
\caption{Distribution of minimum angular distances between the 218 annuli
and the centers of the BATSE error circles.  The angular distances are expressed in
number of sigma for the BATSE error circle radii, i.e.
d/r$_{1\sigma}$.  \label{Fig. 7}}
\end{figure}

\begin{figure}
\plotone{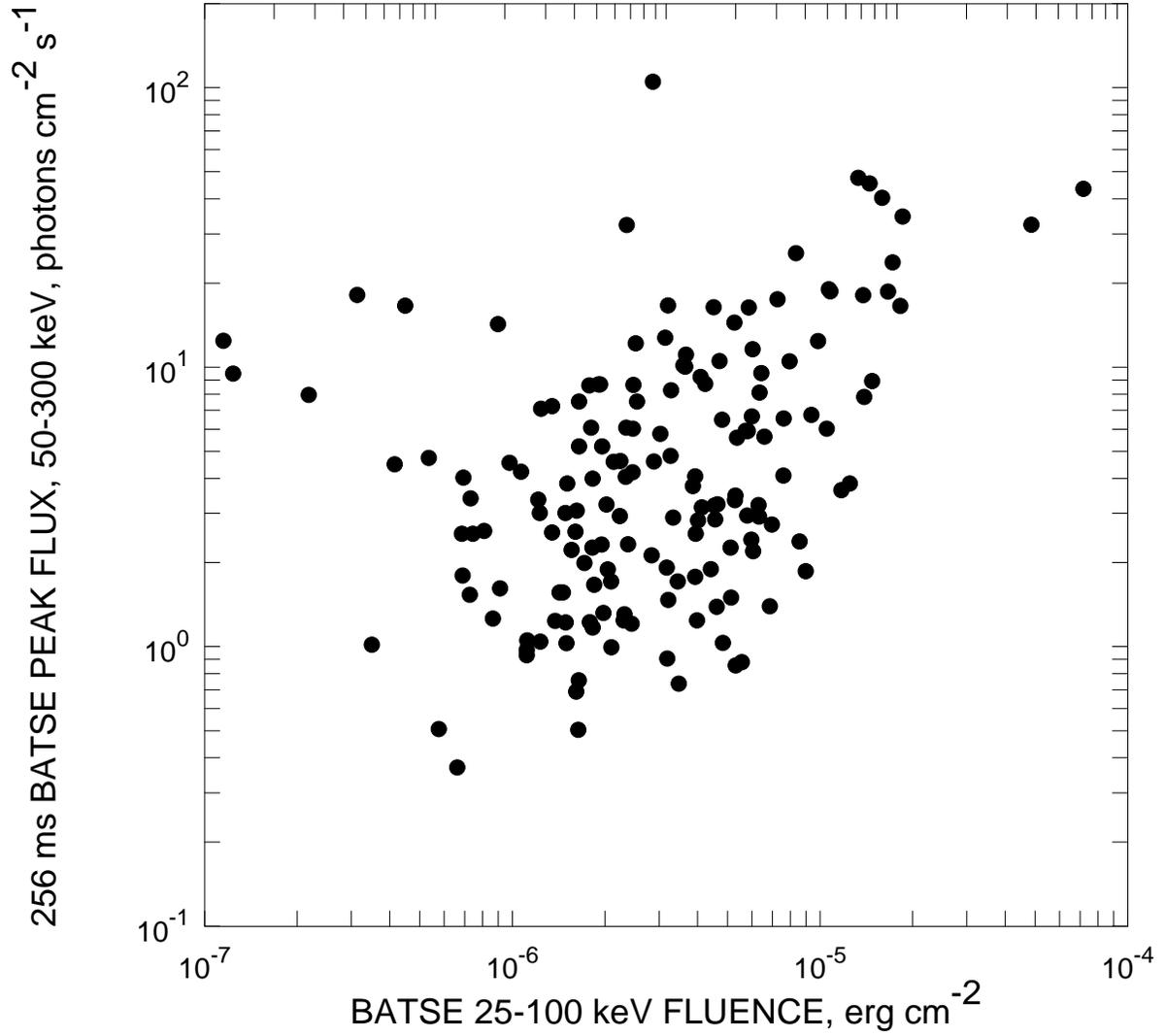}
\caption{Peak fluxes (measured over 256 ms, 50-300 keV) and 25-100 keV fluences of
162 of the bursts in this catalog.  No entries are given in the 3B catalog for 56
of the bursts.  \label{Fig. 8}}
\end{figure}

\begin{figure}
\plotone{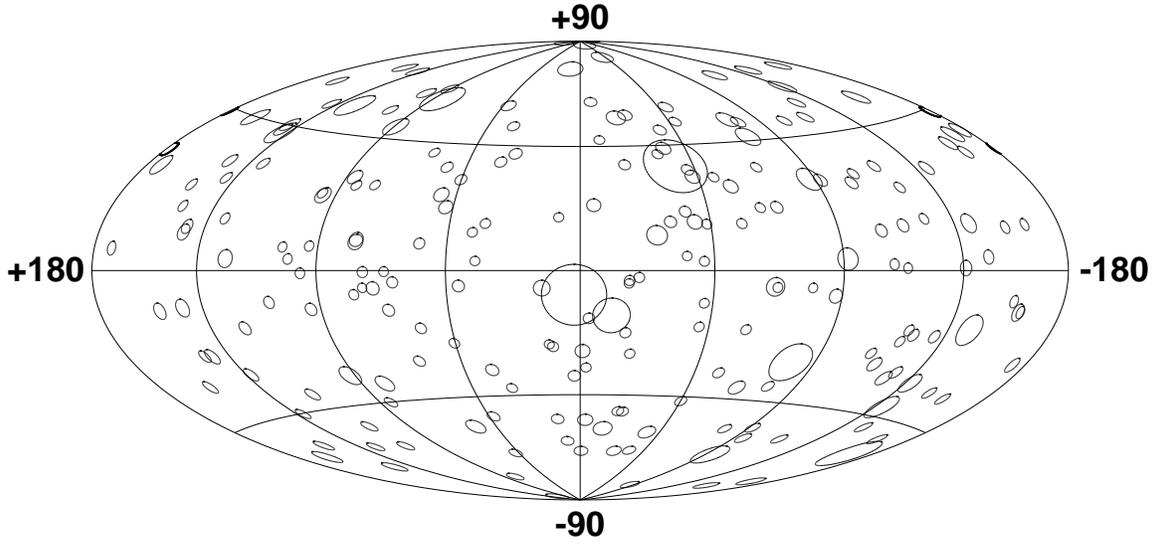}
\figcaption{Distribution of 218 BATSE 3B error circles in galactic coordinates.  
\label{Fig. 9}}
\end{figure}

\begin{figure}
\plotone{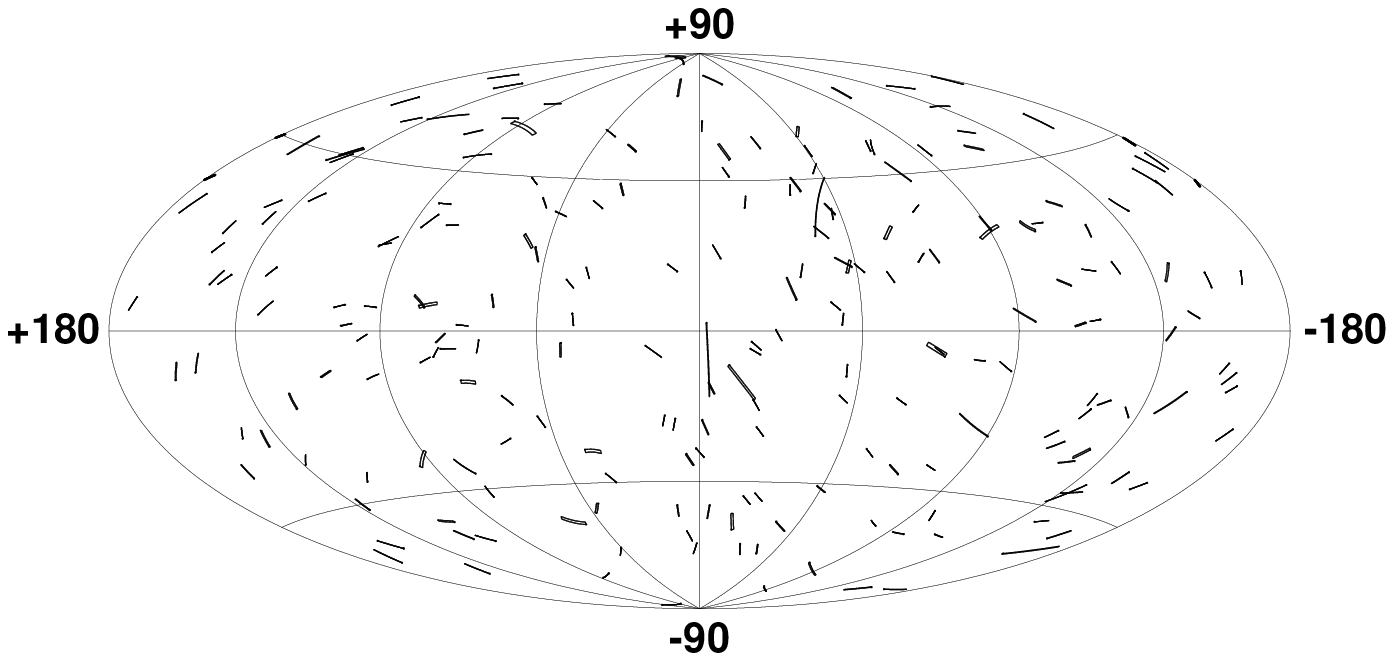}
\figcaption{Distribution of 218 \it Ulysses \rm/BATSE error circle/annulus intersections. 
\label{Fig. 10}}
\end{figure}


\clearpage

\begin{references}
\reference{} Briggs, M., Pendleton, G., Brainerd, J., Connaughton, V., Kippen, 
R. M., Meegan, C., and Hurley, K. 1998a, in
Gamma-Ray Bursts: 4th Huntsville Symposium, eds. C. Meegan, R. Preece, and
T. Koshut, (AIP: New York), p. 104
\reference{} Briggs, M., et al. 1998b, \apjs, in press 
\reference{} Galama, T., et al. 1997, \apj, 486, L5
\reference{} Hurley, K. 1994, in Gamma-Ray Bursts - Second Workshop, eds.
J. Fishman, J. Brainerd, and K. Hurley, AIP Conference Proceedings 307
(AIP: New York), 682
\reference{} Hurley, K., et al. 1992, Astron. Astrophys. Suppl. Ser., 92(2), 401
\reference{} Hurley, K., et al. 1996, \baas, 28(4), 1408
\reference{} Hurley, K., et al. 1997a, \apj, 479, L113
\reference{} Hurley, K., et al. 1997b, Proc. 25th ICRC, Durban, South Africa, OG 2.2.6, 45
\reference{} Hurley, K., et al. 1997c, \apj, 485, L1
\reference{} Hurley, K., et al. 1998a, in
Gamma-Ray Bursts: 4th Huntsville Symposium, eds. C. Meegan, R. Preece, and
T. Koshut, (AIP: New York), p. 50
\reference{} Hurley, K., et al. 1998b, in The Active X-Ray Sky: Results from
BeppoSAX and Rossi-XTE, eds. L. Scarsi, H. Bradt, P. Giommi, and F. Fiore, Nucl. Phys.
B (Proceedings Supplement), 69(1-3), p.660
\reference{} Kippen, M., Hurley, K., and Pendleton, G. 1998, in
Gamma-Ray Bursts: 4th Huntsville Symposium, eds. C. Meegan, R. Preece, and
T. Koshut, (AIP: New York), p. 114
\reference{} Kolatt, T., and Piran, T. 1996, \apj, 467, L41
\reference{} Kulkarni, S., et al. 1998, \nat, 393, 35
\reference{} Laros, J., et al., 1997, \apjs, 110, 157
\reference{} Laros, J., et al., 1998, \apjs, accepted
\reference{} Meegan, C., et al., 1996, \apjs, 106, 45
\reference{} Metzger, M., et al., 1997, \nat, 387, 878
\reference{} Paczynski, B. 1991, Acta Astronomica, 41, 257
\reference{} Schartel, N., Andernach, H., and Greiner, J. 1997, \aap, 323, 659
\reference{} Struble, M., and Rood, H. 1997, \apj, 490, 109
\reference{} Tokanai, F., et al. 1997, \pasj, 49, 207
\end{references}
\end{document}